\documentclass[a4paper]{article}

\usepackage[english]{babel}
\usepackage[utf8x]{inputenc}
\usepackage{amsmath}
\usepackage{graphicx}
\usepackage[colorinlistoftodos]{todonotes}
\usepackage[letterpaper, margin=1in]{geometry}
\usepackage{amsfonts}
\usepackage{mathabx}
\usepackage{algpseudocode}
\usepackage{algorithm}
\usepackage{longtable}
\usepackage{caption}

\title{Investigating the Power of Circuits with $MOD_6$ Gates}
\author{Daniel Saunders}

\begin{document}
\maketitle

\section{Introduction}

We consider the power of Boolean circuits with MOD$_{6}$ gates. First, we introduce a few basic notions of computational complexity, and describe the standard models with which we study the complexity of problems. We then define the model of Boolean circuits, equate a restricted class of circuits with an algebraic model, and present some results from working with this algebra. 

\section{Computational Complexity}

Computational complexity is the study of problems in terms of the resources needed to compute their answers. When a problem is posed, we wish to find a decision procedure that is most efficient in terms of certain constraints (e.g., time or space) that will correctly solve the problem each time.

Depending on the amount of time or space we have access to, or the model on which we are attempting to solve the problem (e.g., a Turing machine, Boolean circuit, or quantum computer), we may or may not be able to come up with such a procedure, and these attempts are what tell us about the fundamental difficulty of the problem in question. In this section, we give some definitions of simple computational concepts and briefly discuss the field of computational complexity. For readers interested in a more in-depth discussion of the field, we recommend one of the many accessible textbooks on the subject \cite{straubing94}, \cite{papdimitriou94}.

\subsection{Languages and Decision Problems}

First, we will give some basic definitions in order that we may talk about computation and what is, or is not, computable. \\

\textbf{Definition 2.1} \textit{A decision problem is a question in some formal system that has a yes or no answer. A decision problem has a Boolean output in $\{0, 1\}$, unlike a functional problem, which can have several-bit solutions.} \\

\textbf{Definition 2.2}\textit{ A formal language is a set of strings of symbols constrained by rules specific to it.} \\

When talking about computational problems, we will often pose them as a decision problem about the membership of the input in a specific language. The use of the term “language” may be somewhat deceptive, as we use it to encompass a wide variety of mathematical problems. For example, the set of strongly-connected graphs form the language $STRONGLY$-$CONNECTED$, and a decision procedure that decides membership in this language would correctly determine whether a graph belongs in the language. We may also extend decision problems to functional problems in a natural way, simply by posing a series of decision problems that will give us a functional answer. For instance, if we wish to add two integers bit-wise, we would pose a series of decision problems, each checking for the expected value of each bit of the result. In this way, it is clear that decision problems are highly extensible and therefore encompass a great deal of complex problems.

\subsection{Turing Machines}

To talk about the nature of computing, it is important that we define a model that may actually carry out any computation we might specify. It must be simple, so that its operation is clear, yet powerful enough to implement any computation we can think of. Below, we define the standard Turing machine model, first proposed by A. M. Turing in 1936. \\

\textbf{Definition 2.3} \textit{A Turing machine $M$ is a model of computation which is defined by the tuple $M = (Q,$ $\Gamma$, $s, F,$ $\delta$), where $Q$ is a finite set of states, $\Gamma$ a finite set of tape alphabet symbols, $s$ the start state, $F$ $\subseteq$ Q the set of final or accepting states, and $\delta$ the transition function, where $\delta$ : $Q$ $\times$ $\Gamma$ $\rightarrow$ $Q$ $\times$ $\Gamma$ $\times$ $\{ L, R \}$.} \\

The Turing machine’s tape is an infinite sequence of cells, each of which contain a letter of the tape alphabet. In the case of circuit complexity, we are only concerned with boolean logic, and so the tape cells will contain only letters selected from $\{0, 1\}$. The tape head is a device that moves along the tape and is able to read or write symbols from the tape alphabet at its current position. The machine functions by maintaining a state from the state set $Q$, where it may transition from one state to another based on the state it is currently in, and the value of the tape alphabet character at which the tape head rests. During any transition, it may also write a new character from the tape alphabet onto its current position, and move left or right on the tape. We may extend this model by changing the way the transition function is defined, but the standard Turing machine model is no less expressive than any non-standard model we care to define.

A fundamental limitation of Turing machines, or any less powerful or equivalent model, is their incapacity to solve certain problems. We call these problems undecidable, and we note that $HALT$, the question of whether a Turing machine $M$ halts on an arbitrary input, is undecidable. So, an algorithm for computing $HALT$ cannot exist, which was shown by A. M. Turing in 1936.

\subsection{Complexity Classes}

We know that Turing machines may compute many complex problems, but we have yet to discuss the difficulty that computing any given problem may have. Considering the time and space resources theoretically needed to compute a given function using a Turing machine will give us a concrete understanding of the difficulty of solving a problem in practice. We will now introduce some well-known general complexity classes. \\

\textbf{Definition 2.4} \textit{The class $P$ is the set of languages that are decidable by a Turing machine using $O(n^{O(1)})$time.} \\

\textbf{Definition 2.5} \textit{The class $L$ is the set of languages that are decidable by a Turing machine using $O(\log(n))$ space.} \\

\textbf{Definition 2.6} \textit{The class $NP$ is the set of languages that are decidable by a nondeterministic Turing machine using $O(n^{O(1)})$ time.} \\

Though these complexity classes are defined in terms of Turing machines, they in fact correlate to a class of problems independent of the model they are computed on. This suggests that these classes are natural and robust, and allows for the study of the fine structure of these classes within different computation models. It has been established that the expressiveness of a Turing machine with a polynomial amount of time is equivalent to that of polynomially large uniform Boolean circuits or Boolean queries using first-order logic with polynomially many blocks of quantifiers. The main goal of complexity theory is to understand the relationship between these complexity classes, and to that end, we use different models in order to study the fine structure of them. One of the most important open questions in computer science is whether $P = NP$ or not. It has been shown by Lipton and Karp that if $NP$ has polynomially sized circuits, $P = NP$. Since we have many tools for proving circuit lower bounds, many lines of research have been in attempting to prove a circuit lower bound on $NP$, but with little success.

\section{Algebra, Logic, and Boolean Circuits}

Now, we introduce three more computation models, those which will help motivate the main question of this paper. \\

\textbf{Definition 3.1}\textit{ A monoid $M$ is a set together with an operation, $* : M \times M \rightarrow M$ such that $*$ is associative, and $M$ contains an identity element $e$ such that $e*m = m*e = m$, $\forall m \in M$}. \\

A monoid is simply a group which does not require inverses, a concept familiar to those who have studied group theory. A particularly important group, for our purposes, is the permutation group on $n$ elements, denoted by S$_n$. This is the group of actions on a set by permutation. There are clearly $n!$ elements of S$_n$, as there are $n!$ permutations of $n$ elements. We will denote an element of S$_n$ in the standard cycle notation; for example, the cycle $(2 3 1)$ sends the element 1 to 2, 2 to 3, and 3 to 1. The product of two cycles is the composition of the two permutations.

\subsection{Algebraic Programs}

Using algebra, we present another model of computation apart from Turing machines. An $M$-program is a sequence of instructions that are simply monoid elements, which allows us to compute a function by computing the product of the sequence of monoid instructions. \\

\textbf{Definition 3.2} \textit{An $M$-program is a sequence of instructions that can be modeled by functions from a given input bit to an element of a monoid. The result of the $M$-program is the product of the monoid elements. So, a program $\Phi$ with n input bits i$_1$, …, i$_n$ and functions f$_m$ : $\{0 , 1\}$ to $M$ can be expressed as a sequence of instructions:} \\

\begin{center}
		$\Phi = (i_{x_1}, f_1)...(i_{x_l}, f_l)$,
\end{center}

\textit{where each term is an instruction to evaluate the function $f_i$ on the $x^{th}_z$ bit. On input a, $\Phi$'s output is given by:} \\
 
 \begin{center}
 		$\Phi (a) = f_1(i_{x_1})...f_l(i_{x_l})$, 
\end{center}
  
\textit{which accepts if and only if $\Phi$($a$) in $M$’ for a given $M$’ in $M$.} \\

So, the M-program generates a sequence of instructions, and then evaluates the word problem for the monoid M. We denote the length of the program by the number of instructions used. The computational power of different monoids and groups is important to us, since the Tsukiji problem can be expressed as a program over a certain group.

\subsection{The Group $G_{72}$}
We are interested in a particular group, denoted by $G_{72}$, and its actions on $\mathbb{Z}_3 \times \mathbb{Z}_2$, which is of interest to our problem. We present some important relations and identities between the five generators of this group (a, b, c, d, e) in Table 1. 

This provides an interesting algebraic foundation for the Tsukiji problem, which is central to our investigation of the power of circuits with $MOD_6$ gates. There is an equivalence, within a polynomial factor, of programs over this group to the class of function computable by depth 3 $MOD_3$ of $MOD_2$ gate circuits (a formal proof is given in \cite{sindelar15}), and so finding a lower bound on the size of this circuit is equivalent to finding a lower bound for the length of programs over this group.

\begin{center}
Table 1: The Group $G_{72}$
\break
\begin{tabular}{ |c|c|c| }

\hline
$a^3 = 1$ & $b^3 = 1$ & $ab = ba$ \\ 
\hline
$c^2 = 1$ & $cac = b$ & $cdc = e$ \\
\hline
$d^2 = 1$ & $dad = a^2$ & $db = bd$ \\
\hline
$e^2 = 1$ & $ebe = b^2 $ & $ea = ae$ \\

\hline
\end{tabular}
\end{center}

\hfill

\subsection{First Order Logic and Complexity}

Here, we briefly introduce the idea of descriptive complexity, the study of computation from the perspective of logic. We can ascertain the difficulty of a problem by investigating what tools are needed to express the problem with logical queries. We present a few results here that characterize certain classes of interest to our problem at hand with their logical counterparts. For further reading on descriptive complexity and proofs of the following claims, consult \cite{immerman99}. \\

\textbf{Definition 3.3} \textit{$FO$ is the class of problems whose solutions can be expressed using first-order logical queries.} \\

We can extend the power of first order logic by adding new quantifiers, and in turn extend the expressiveness of $FO$. We are mainly interested in the modular counting quantifier and the majority quantifier, which are relevant to the Tsukiji problem. The modular counting quantifier, for counting mod $p$, is true if and only if there are exactly 0 mod $p$ values for $x_i$ such that $\phi$ is true. Similarly, the majority quantifier is true if and only if there are a majority of values of $x_i$ such that $\phi$ is true. Adding these quantifiers, we define extensions of $FO$ as follows: \\

\textbf{Definition 3.4} \textit{$FO[MOD_p]$ is the class of problems whose solutions can be expressed using first-order logical queries with counting quantifiers modulo p.} \\

\textbf{Definition 3.5:} \textit{$FO[Maj]$ is the class of problems whose solutions can be expressed using first-order logical queries with majority quantifiers}. \\

It can be shown that the class $FO[MOD_p]$ is equivalent in power to the circuit class $AC^{0}[p]$, and the class $FO(Maj)$ is equivalent in power to the circuit class $TC^{0}$, both of which we shall define in the next section.

\section{Boolean Circuits and Complexity}

We now introduce our model of interest, Boolean circuits, which are collections of interconnected gates which perform basic computations. This is different from such an abstraction as the Turing model of computation, since it allows us to look very closely at low-level computation, and find lower bounds on functions of little complexity. 

\subsection{Definitions}

First, we define some basic components of the circuit model. \\

\textbf{Definition 4.1} \textit{A gate is a component of a circuit that performs a basic computational function}. \\

In the standard circuit model, we consider only the gate functions $\land$, $\lor$, and $\lnot$, which constitute the standard basis $\hat{B}$, but here we consider the added expressiveness of circuits with counting (modulo $p$) and majority gates. Adding these gates constitutes the circuit classes $AC^{0}[p]$ and $TC^{0}$, respectively. In particular, the Tsukiji problem requires $AND$, $MOD_{3}$, and $MOD_{2}$ gates. 

Now, a circuit is a collection of gates, namely, a set of input gates, output gates, and other computation gates which are connected via wires. In particular, the input gates are typically connected to some collection of computation gates, and eventually, each path from an input gate will terminate in an output gate. In this way, the circuit computes a function on its input bits and outputs the bit or bit string stored in the output gates. Circuits are acyclic, and we therefore avoid any wire connections from any computation gate to any input gate or ancestral gates. This allows us to avoid issues with timing, changing output values, and circuits that evolve over time. \\
    
\textbf{Definition 4.2} \textit{A circuit can be modeled by the tuple $C = (G, W, I, O)$. Here $G$ is a set of gates and $W$ a set of wires connecting the gates, which together form a directed acyclic graph. $I$ is the set of gates corresponding to input bits, and $O$ the set of output gates. The value of a gate is the output of the function it computes on the wires that connect into it. The output of a gate is passed on the output wires. The output of the circuit is the output value(s) of the output gate(s)}. \\

Oftentimes, a circuit will simply compute a decision problem, which restricts the circuit to producing one bit, but this construction can easily be extended to functional problems. In order to measure the complexity of a given circuit, we consider its size (the number of gates or wires) and the depth (the longest path from an input to output gate). We also consider the fan-in (the maximum number of wires used for input to a gate) and the fan-out (the maximum number of wires used for output from a gate) of the circuit in question. The complexity measures of size and depth loosely correspond to the measures of space and time, respectively, which result from the Turing model of computation. We say that a circuit has unbounded fan-in or fan-out to say that we may use an unlimited amount of wires as input or as output, respectively. 

As we have defined circuits, they may only compute a function with an input of a given length. Turing machines, on the other hand, may compute a function given an input of arbitrary length. In order to extend our model to have this same capacity, and to be able to compare the two models, we must introduce the notion of circuit families. \\

\textbf{Definition 4.3} \textit{A circuit family which computes a function $f$ is a collection of circuits $C_{i}$ such that for every $i \in \mathbb{N}$, $C_{i}$ computes $f$ on inputs of length $i$.} \\

We can now describe the asymptotic complexity of a function in terms of the size and depth of the circuit family that computes it. Furthermore, augmenting or restricting our basis functions and the fan-in and fan-out of the circuit will determine the expressiveness of the circuit classes we care to define.

\subsection{Bounds for General Boolean Functions}

We may now introduce a few well-established complexity bounds on certain Boolean functions of interest to our problem. The work of Shannon and Lupanov established asymptotic lower and upper bounds on most Boolean functions on $n$ variables. \\

\textbf{Theorem 4.1} \textit{For $\epsilon > 0$, the ratio of $n$-ary boolean functions computable by circuits over $\hat{B} = B_{0} \cup B_{1} \cup B_{2}$ with $(1 − \epsilon)\cdot2^{n}/n$ gates approaches 0 as $n \rightarrow \infty$. In other words, for large $n$, most Boolean functions have size $\Omega (2^{n}/n)$. }\\

\textbf{Theorem 4.2} \textit{Every $n$-ary Boolean function can be computed by circuits with $2^n / n + o(2^n / n)$ gates over the basis $\{ 0, 1, \oplus, \land \}$.} \\

Combining these two theorems, it is easy to see that for large $n$, most Boolean functions have complexity $\Theta(2^{n}/n)$. However, we are interested in studying functions of complexity $O(n^{O(1)})$, which are considered feasible to implement in Boolean circuits.

\subsection{Uniformity}

It is important to mention how individual circuits are constructed, given any circuit family. For example, the halting problem is known to be uncomputable in the Turing paradigm, but we may define a circuit family which solves the unary version of this problem. On inputs of length $n$, if Turing machine $n$ halts on itself, the circuit is simply the the constant gate 1, otherwise, it is the constant gate 0. This is obviously a well-defined circuit family, but is not constructible, since determining how to build the individual circuits would give us a decision procedure for the halting problem. 

To mitigate this conceptual problem, we introduce the notion of uniformity. A uniform circuit family is one whose gates may be described by a Turing machine, constrained with polynomial time or logarithmic space. This allows us to prove equivalences between Turing classes and circuit classes. We may also show equivalences by augmenting Turing machines with an “advice” tape which varies with input size, thus making it a non-uniform machine, which could allow it to have the same non-uniform power that circuits may have. Clearly, the circuit family that decides the halting problem is a non-uniform circuit class, as it cannot be constructed by a Turing machine given even infinite time or space resources.

\subsection{Circuit Complexity Classes}

We now define a number of complexity classes in which to place problems computable by circuits. We are mostly interested in circuits with no more than polynomial size or polylogarithmic depth as a function of the input. These highly constrained classes of circuits cannot easily simulate other computational models, and therefore the basis from which we may select our gate functions is extremely important in determining their power. We shall assume that the following classes we define are all logarithmic space-uniform, and can therefore be described (or built) by a Turing machine with access to logarithmic space. Furthermore, we assume that, unless stated otherwise, that each circuit class we define may select gates from the standard basis $\hat{B} = \{ \land, \lor, \lnot \}$. \\

\textbf{Definition 4.4} \textit{$AC^i$ is the set of all languages that are recognizable by polynomially sized circuits of unbounded fan-in gates over $\hat{B}$ and $O(\log^i(n))$ depth.} \\

\textbf{Definition 4.5} \textit{$NC^i$ is the set of all languages that are recognizable by polynomially sized circuits of bounded fan-in gates over $\hat{B}$ and $O(\log^i(n))$ depth}. \\

\textbf{Definition 4.6} \textit{$AC^i[m]$ is the set of all languages that are recognizable by polynomially sized circuits of bounded fan-in gates over $\hat{B} \cup \{ MOD_m \}$ gates and $O(\log^i(n))$ depth. The union of $AC^i[m]$ for all m $\in \mathbb{N}$ is denoted as $ACC^i$, known as $AC^i$ with counters.} \\

\textbf{Definition 4.7} \textit{$CC^i[m]$ is the set of all languages that are recognizable by polynomially sized circuits of unbounded fan-in $MOD_m$ gates and binary $\land$ and $\lor$ gates and $O(\log^i(n))$ depth.} \\

\textbf{Definition 4.8} \textit{$TC^i$ is the set of all languages that are recognizable by polynomially sized circuits of bounded fan-in gates over $\hat{B} \cup \{ MAJ \}$ gates and $O(\log^i(n))$ depth}. \\

Now, a $MOD_m$ gate returns 1 if and only if the bit-sum of its inputs is equal to 0 modulo m, and returns 0 otherwise. Similarly, a $MAJ$ gate returns a 1 if and only if half or more of its input bits evaluate to\textit{} 1. Now that we have these definitions of uniform circuit classes, we will mention some of their important relationships to standard complexity classes. \\

\textbf{Theorem 4.3} \textit{$NC^1 \subseteq L \subseteq NL \subseteq AC^1$.} \\

\textbf{Proof} We can show that $NC^1 \subseteq L$ by constructing an algorithm that evaluates $NC^1$ circuits in logspace, and thus, a Turing machine with logarithmic space may simulate any $NC^1$ circuit. We may do this with a simple recursive algorithm that uses boolean operators. $NC^1$ circuits have only bounded $\land$, $\lor$, and $\lnot$ gates (where we may push the $\lnot$ gates to the bottom and therefore eliminate them from the circuit). The algorithm works as follows. We start at the output gate. For the recursive step: If the gate we are at is an $\lor$ gate, return the logical $\lor$ of the values of the algorithm at the right parent and the left parent. Similarly, if the gate we are at is an $\land$ gate, return the logical $\land$ of the values of the algorithm at the right parent and the left parent. If the gate we are at is an input gate, return the value of the gate. Clearly, this will evaluate the circuit as it was designed to be, and requires only logarithmic space to keep pointers of where it is in the recursion, which has no more than logarithmic depth.

Since $NC^1$ has only $O(\log(n))$ depth, we need only keep track of $O(\log(n))$ bits of pointers (since each node has some bounded number of children). And so, for logspace uniform circuits, there exists a Turing machine which can describe the circuit using logarithmic space. We can use the machine to construct such circuits, and so we may evaluate any circuit in $NC^1$ using logarithmic space, implying that $NC^1$ is contained in $L$.

$L \subseteq NL$ is trivially true, as for any $L$ machine, we may construct an $NL$ machine that behaves identically by copying the $L$ machine and not including any nondeterministic transitions. So this $L$ machine is simulated by our copycat $NL$ machine.

We may show that $NL \subseteq AC^1$ by constructing an $AC^1$ circuit which solves an $NL$-complete problem. We will use $REACH$, the problem of whether a graph $G$ contains a path from node $s$ to node $t$, as our $NL$-complete problem. Consider the predicate $PATH(s, t, l)$ which is true if and only if there exists a path from node $s$ to node $t$ of length less than or equal to $l$.  We create a root node that is equivalent to $PATH(s, t, n)$, where $n$ is the total number of nodes in the graph, and therefore the maximum length of ant path in the graph. We wish to see if a midpoint exists; we make this root node an unbounded $\lor$ gate on $PATH$-$THROUGH(s, m_1, t, n)$ for every node m where $PATH$-$THROUGH(s, m, t, n)$ means there is a path from $s$ to $t$ with midpoint $m$. These $PATH$-$THROUGH(s, m_1, t, n)$ nodes can be expressed as $\land$ gates on $PATH(s, m_1, n/2)$ and $PATH(m_1, t, n/2)$. We can then define these as $\lor$ gates on all $m_i$ and then $\land$ gates as above, and so on in this alternating fashion, until we reach nodes of the form $PATH(x, y, 1)$, which we may treat as input nodes based on whether there is an edge from $x$ to $y$ or $x = y$. Clearly this is still $O(\log(n))$ depth, since we halve $n$ every other step until it is 1, and it is polynomially size because at each depth we only consider ordered pairs or ordered triples of nodes, since each gate is asking whether there is a path from node $a$ to $b$ or from $a$ to $b$ through $c$. Clearly, there can be at most $n^3$ nodes at each level, and there are $O(\log(n))$ levels, and so the total size of this construction is $O(\log(n) \cdot n^3)$, or $O(n^4)$. \\

\subsection{Lower Bounds}

Here we state some lower bound results concerning circuit and logic complexity classes, without proof. Please refer to \cite{vollmer99} for detailed proofs. \\

\textbf{Theorem 4.5 (Furst-Saxe-Sipser)} \textit{The parity function, or counting modulo 2, is not contained in $AC^0 = FO$. Equivalently, we say that the parity function cannot be expressed by a circuit with constant depth, polynomial size, and unbounded fan-in, with gates selected from the standard basis $\hat{B} = \{ \land, \lor, \lnot \}$.} \\

\textbf{Theorem 4.5 (Smolensky's Theorem)} \textit{For distinct primes $p$ and $q$, counting mod $q$ cannot be performed by circuits of polynomial size, constant depth, unbounded fan-in, and gates selected from $\hat{B} \cup MOD_p$. Equivalently, $AC^0[q] \nsubseteq AC^0[p]$, and vice verses.} \\

Smolensky showed that each prime counting class cannot capture the expressiveness of any relatively prime counting classes. This, however, leaves open the question of the power of circuits which include counting gates of two relatively prime integers. As an instance of this problem, we consider the work by Tsukiji on the power of gates with both $MOD_2$ and $MOD_3$ gates, and ask whether or not this class of circuits has some unexpected computational power.

\section{The Constant Degree Hypothesis}

Our main avenue of investigation was into the Constant Degree Hypothesis, a conjecture on the expressiveness of sums of polynomials and restricted circuit classes. We present another model for understanding the computational power of these circuits we have mentioned above, using multilinear polynomials. Following are some definitions and theorems on the subject, and, unless stated otherwise, the theorems presented here are Barrington's. \\

\textbf{Definition 5.1} \textit{A multilinear polynomial is a polynomial over $F[x_1, x_2, ..., x_n]$ that is linear in all of its variables, i.e. it contains no terms of the form $x{^p}{_i}$ for $p > 1$. Its degree is the degree of its maximum term, which is equal to the number of variables in the term.} \\

\textbf{Definition 5.2} \textit{A linear form, over a field $F$, on $n$ variables, $x_1, x_2, ..., x_n$, is a polynomial represented by $\sum_{i=1}^{n} a_ix_i + a_0$, with each $a_i \in F$. }\\

\textbf{Definition 5.3} \textit{A quadratic form, over a field $F$, on $n$ variables, $x_1, x_2, ..., x_n$, is a polynomial represented by $\sum_{i=1}^{n} \sum_{j=1}^{n} a_{i, j}x_ix_j + \sum_{i=1}^{n} a_ix_i + a_0$, with each $a_i \in F$.} \\

\textbf{Definition 5.4}\textit{ A linear character, over a field $F$, is a function from strings of length n to $F$, given by $\omega^l$, where $l$ is a linear form and $\omega$ is a generator for $F$.} \\

\textbf{Definition 5.5} \textit{A quadratic character, over a field $F$, is a function from strings of length n to $F$, given by $\omega^q$, where $q$ is a quadratic form and $\omega$ is a generator for $F$.} \\

For our purposes, we will only consider forms over the field $\mathbb{Z}_2$ and characters over the field $\mathbb{Z}_3$, and thus, $\omega =$ 2, as 2 acts as a generator for $\mathbb{Z}_3$. By summing together these characters, we may compute functions by mapping a set of $n$ input bits to an output integer selected from $\mathbb{Z}_3$, computed arithmetically. We consider the number of characters in said sums as a measure of complexity. \\

\textbf{Definition 5.6} \textit{The support of a function is the number of input strings on which the function evaluates to a non-zero number.} \\

\textbf{Definition 5.7} \textit{The $n$-weight of a function is the minimum number of degree $n$ characters needed whose sum is equal to the function. The 1-weight of a function is the number of linear characters needed to compute it, and the 2-weight is the number of quadratic characters needed}. \\

We note that the following proofs are taken from the Sindelar paper \cite{sindelar15}, and have been modified in order to improve readability or to narrow their scope. \\

\textbf{Theorem 5.1} \textit{The complexity of a function under the following models is the same up to within a polynomial factor:} 

\begin{enumerate}
	\item \textit{Size of depth 2 circuits formed by a $MOD_3$ gate of $MOD_2$ gates.}
    \item \textit{Sums of linear characters over $\mathbb{Z}_3$ with forms over $\mathbb{Z}_2$.} 
    \item \textit{Lengths of programs over $S_3$}
\end{enumerate}

\textbf{Proof (1 $\Rightarrow$ 2)} The sums of $s$ linear characters may be simulated by circuits as follows. We connect a $MOD_2$ gate to each linear form and the constant gate 1, if it is included in the form, in the exponent of each linear character, which will compute its parity. We connect an additional constant gate to these $MOD_2$ gates, which will flip the parity bit, and so the output of this gate with be 0 if the parity of the linear form is even, and 1 otherwise. We couple each $MOD_2$ gate with a constant $MOD_2$ gate (which will always output 1), and so if the form evaluates to 1, the bit-sum of these two gates is 2, and otherwise, 1. We connect all of these to a single $MOD_3$ gate, which will then compute the sum of all $s$ linear characters modulo 3. So we have constructed a depth 2 circuit of a $MOD_3$ gate of $MOD_2$ gates, using only $O(s)$ gates to do so.

Similarly, we may simulate an arbitrary depth 2 circuit of a $MOD_3$ gate of $MOD_2$ gates with a sum of linear characters over $\mathbb{Z}_3$ with forms over $\mathbb{Z}_2$ as follows. Suppose the depth 2 circuit contains s $MOD_2$ gates. For each $MOD_2$ gate, we define a linear form $l$ in which $x_i$ is present if the input $x_i$ is connected to the gate in question. We then define a linear character, $2^l$, for each of these forms, and, taking the sum, we are done. So we are able to simulate arbitrary circuits of this form with only $s$ linear characters. 

Here, we do not prove the equivalence, within a polynomial factor, of the complexity (given by length) of programs over $S_3$ to these models. The reader may consult \cite{sindelar15} for a formal proof. \\

A similar result holds for sums of quadratic characters over $\mathbb{Z}_3$ with forms over $\mathbb{Z}_2$. \\

\textbf{Theorem 5.2} \textit{The complexity of a function under the following models is the same up to within a polynomial factor:}

\begin{enumerate}
	\item \textit{Size of depth 3 circuits formed by a $MOD_3$ gate of $MOD_2$ gates of binary $\land$ gates.}
    \item \textit{Sums of quadratic characters over $\mathbb{Z}_3$ with forms over $\mathbb{Z}_2$.}
    \item \textit{Lengths of programs over $G_{72}$.}
\end{enumerate}

\textbf{Proof} This proof is similar to that of Theorem 5.1. We may simulate sums of quadratic characters over $\mathbb{Z}_3$ with forms over $\mathbb{Z}_2$ by simulating the quadratic terms with binary $\land$ gates (e.g., $x_1x_2$ = 1 if and only if $x_1 \land x_2 = 1$), and the rest of the construction follows that of Theorem 5.1.
	
Likewise, the depth 3 circuits of a $MOD_3$ gate of $MOD_2$ gates of binary $\land$ gates may be simulated by sums of quadratic character over $\mathbb{Z}_3$ with forms over $\mathbb{Z}_2$ by constructing the sums analogously to those of Theorem 5.1, but with the added simulation of the $\land$ gates by the quadratic terms in the forms belonging to the quadratic characters. 

Again, the reader may consult $\cite{sindelar15}$ for a formal proof of the equivalence, within a polynomial factor, of the complexity of programs over $G_{72}$ to these two models. \\

\textbf{Conjecture 5.1 (The Constant Degree Hypothesis)} \textit{$2^{n^{\Omega (1)}}$ constant degree characters are required to sum to the $AND_n$ function over a finite field $F$.} \\

Barrington showed that a programs over $S_3$ require exponential length to perform $AND$, and together with the result that the complexity of programs over $S_3$ is equivalent to the number of characters required for a sum (up to a linear factor), this tells us that an exponential number of linear characters are required for $AND$. We provide an alternate proof (due to Tsukiji), via a probabilistic argument, which shows that an exponential number of linear characters are needed to sum to $AND$ over any finite field $F$, not just $\mathbb{Z}_3$. \\

\textbf{Theorem 5.3 (Lower Bound for AND with Linear Characters} \textit{Over a finite field of order k, $O((k/(k+1))^n)$ linear characters are required to sum to $AND_n$.} \\

\textbf{Proof} Let $AND_n$ be the sum of linear characters $\sum c_\alpha \omega ^{\alpha}$, and let $r(x)$ be a random linear form over $\mathbb{Z}_k$. Then $\omega ^{r(x)} AND_n = \omega ^{r(x)} \sum c_\alpha \omega ^{\alpha} = \sum c_\alpha \omega ^{\alpha + r(x)}$. Clearly, $\alpha + r$ is a linear form, as both are linear forms themselves. If there exists some $x_i \neq 1$, then $\omega ^{r(x)} AND_n = \omega ^{r(1)}$, where $r(1)$ is the linear form $r(x)$ with all inputs as 1's. So, $\omega ^{r(1)}AND_n = \omega ^{r(1)}(\prod_{i=1}^{n}x_i)$ and must have degree n. Since this is equal to $\sum c_\alpha \omega ^{\alpha + r(x)}$, this must also have degree n, which implies that at least one of the terms in the sum has degree n. But $\sum c_\alpha \omega ^{(\alpha + r)(x)}$ has degree n only when $(\alpha + r)(x)$ does, which only occurs when, for each $x_i$, the coefficient of $x_i$ in $r$ is not the additive inverse of the coefficient of $x_i$ in $\alpha$. For any given $x_i$, randomly choosing an element that is not the additive inverse occurs with probability $(k-1)/k$ over a field $\mathbb{Z}_k$, and since there are $n$ terms in the sum, the total probability of a random character added to $\alpha$ having degree n is $((k-1)/k)^n$. So there must be at least $(k/(k-1))^n$ linear characters in order to guarantee there is a term with degree $n$. We conclude that we need at least $(k/(k+1))^n$ linear characters to sum to $AND_n$ over $\mathbb{Z}_k$. \\

\textbf{Theorem 5.4} \textit{The 2-weight of $AND_n$ is $\leq$ $2^{n/2}$.} \\

\textbf{Proof} This is clear from the following product of quadratic characters:

\begin{equation*}
  \prod_{i=1}^{n/2} (2^1 + 2^{x_{2k-1}x_{2k}}) = AND_n
\end{equation*}

This product is the sum of $2^{n/2}$ quadratic characters, because each multiplication of polynomials of two terms at most doubles the number of terms in the product. This product is equal to $AND_n$, since if all the bits evaluate to 0, then each term in the product takes the form $2^1 + 2^{1*1} = 2 + 2 = 1$ (mod 3), and so the total product is equal to 1. Further, if any bit evaluates to 0, there there is a term in the product such that $2^1 + 2^0 = 2 + 1 = 0$ (mod 3), and so the total product is equal to 0. So the 2-weight of $AND_n$ is less than, or equal to, $2^{n/2}$.

We wish to know whether or not this is the optimal way to product the $AND_n$ function. If it is, it would prove that $AND_n$ is not contained in polynomially sized circuits of this form. We conjecture that this is the case, as we have not been able to find a more optimal way to produce this function. \\

\textbf{Conjecture 5.2 (Tsukiji)} \textit{The 2-weight of $AND_n$ is exactly $2^{n/2}$.}

\subsection{Witt Rank and Decomposition}

One way of classifying the different quadratic characters is to decompose them into linearly independent terms. By doing this, we may draw isomorphisms between characters that seem different, but are really the same with respect to a change of basis that maps linear terms to linear terms. A character's Witt decomposition allows us to consider such sets, and its Witt rank allows us to compare different quadratic characters. \\

\textbf{Definition 5.8} \textit{The Witt decomposition of a quadratic form $q$ is an expression $l_1l_2 + l_3l_4 + ... + l_{2r−1}l_{2r}+l_0$ where the $l_i$’s are linearly independent linear forms. The Witt rank is the number $r$ in the decomposition.} \\

\textbf{Definition 5.9} \textit{For any $n$, there are $2n$ + 2 unique Witt decompositions. By change of basis, any quadratic form $q$ may be written in the one of the quadratic forms $\{0, 1, x_1, x_1 + 1, x_1x_2, x_1x_2 + 1, ..., x_1x_2 + ... + x_{n-1}x_n, x_1x_2 + ... + x_{n-1}x_n + 1\}$, in what we call its Witt Normal Form.} \\

We note again that the following proofs are taken from the Sindelar paper \cite{sindelar15}, some of which are slightly modified to increase readability or accuracy. \\

\textbf{Theorem 5.5} \textit{The Witt rank of a quadratic form is unique.} \\

\textbf{Proof} We must reference a few facts that are proven later in this paper in order to prove this theorem. In Theorem 5.8 we show that any two quadratic forms that differ only by linear terms have the same Witt rank. For an arbitrary quadratic form, consider the family of all quadratic forms that have the same pure quadratic part, i.e., the same quadratic terms. One of these forms has support that is a function of the Witt rank, by Theorem 5.11, and therefore must have a unique Witt rank since support is clearly unique. So there are two functions with unique Witt rank and support in the family, $q_1$ and $q_2$, with Witt rank $r_1$, $r_2$, and support $2^{n-1} + 2^{n-r_1-1}, 2^{n-1}+2^{n-r_2-1}$, respectively. Since $q_1$, $q_2$ are in the same family, they must differ by a linear term. This linear character is non-constant, since it may not be 0, and if it were 1, $2^{n-1}+2^{n-r_1-1}=2^{n-1}+2^{n-r_2-1}$, which is never true for any $r_1$, $r_2$ $\in \mathbb{N}$. There cannot be a difference in support $2^{n-r_1-1}-2^{n-r_2-1} < 2^{n+1}$, since adding a non-constant linear term can only change the support by $2^{n-1}$ if the linear is not in the quadratic form, or by 0 if the linear is already in the form. \\ 

Now that we have defined Witt rank, we wish to see how a form's rank informs some of its basic properties. First, we shall show how the Witt rank provides a bound on the minimum number of linear characters needed to construct a function. \\

\textbf{Theorem 5.6} \textit{Any quadratic character of Witt rank $r$ is the sum of at most $4^r$ linear characters.} \\

\textbf{Proof} We sketch the Witt decomposition algorithm below, and we use it to write $2^q$ as $2^{l_1l_2} + ... + 2^{l_{2r-1}l_{2r}} + 2^{l_0} = 2^{l_0}\prod_{i=1}^{2r} 2^{l_{i-1}l_i}$. We note that we may write any $2^{l_1l_2}$ as $2^1 + 2^{l_1+1} + 2^{l_2+1} + 2^{l_1+l_2}$. To see this, consider the table below that depicts all the possible values of the linear terms $l_i$. Clearly, we have shown that the sums of these linear characters and the original quadratic character are equivalent, and so we may write any term in the product as the sum of four linear characters, and the product of $r$ polynomials of at most 4 terms is at most $4^r$. \\

\begin{algorithm}
    \caption{Witt Decomposition}
    
    \begin{algorithmic}[1]
        \Require $q$ is a quadratic form
        \Ensure the $l_i$'s are the terms of the Witt decomposition
        \State $r = 0$;
        \State $i = 1$;
        \While {q is not linear} 
            \If {$x_ix_j$ appears in $q$ for some $j$ \textbf{then}} 
                \State $j $ = minimum such $j$;
                \State $r++$;
                \State choose $l_{2r−1}$ and $l_{2r}$ such that $q-l_{2r−1}l_{2r}$ is free of $x_i$, $x_j$;
                \State $q = q - l_{2r−1}l_{2r}$;
            \EndIf
            \State $i++$;
        \EndWhile
        \State $l_0 = q$;
    \end{algorithmic}
\end{algorithm}

\begin{center}
Table 2: Writing a Quadratic Character as a Sum of Linear Characters \break \\
\begin{tabular}{ |c|c|c|c|c|c|c|c| }
\hline
$l_1$ & $l_2$ & $2^{l_1l_2}$ & $2^{l_1+1}$ & $2^{l_2+1}$ & $2^1$ & $2^{l_1+l_2}$ & $\sum$ \\
 \hline
 0 & 0 & 1 & 2 & 2 & 2 & 1 & 1 \\ 
 0 & 1 & 1 & 2 & 1 & 2 & 2 & 1 \\ 
 1 & 0 & 1 & 1 & 2 & 2 & 2 & 1 \\ 
 1 & 1 & 2 & 1 & 1 & 2 & 1 & 2 \\ 
 \hline
\end{tabular}
\end{center}

\hfill

\textbf{Theorem 5.7} \textit{Algorithm 1 correctly produces the Witt decomposition of any quadratic form $q$.} \\

\textbf{Proof} On each iteration, this algorithm will take two linear terms and create two independent basis elements, and removes all instances of them from the form. So, on the next iteration, the new basis elements must be independent since they cannot contain the linear terms that have been removed. \\

\textbf{Theorem 5.8} \textit{Quadratic forms with the same pure quadratic part, i.e., that only differ in linear or constant terms, have the same Witt rank.} \\

\textbf{Proof} The idea is to select $l_{2r-1}$ and $l_{2r}$ such that the quadratic form $q - l_{2r-1}l_{2r}$ is free of some $x_i$, $x_j$. We accomplish this by letting $l_{2r-1}$ be the sum of linear terms which include $x_i$ and $x_m$ such that, for all $x_m$ for which $x_ix_m$ is a quadratic term in $q$. Analogously, we let $l_{2r}$ be the sum of linear terms which include $x_j$ and $x_n$ such that $x_jx_n$ is a quadratic term in $q$. We also add the constant 1 to $l_{2r-1}$ if the linear term $x_i$ is in $q$, and add the constant 1 to $l_{2r}$ if the linear term $x_j$ is in $q$. So if the pure quadratic part of $q_1$ and $q_2$ are the same, the only difference in the first iteration is the possible inclusion of the constant 1. This means that the difference between the two sets of $l_{2r-1}$, $l_{2r}$ will be the sum of linear characters, the product of $l_{2r+1}$ with the constant and the product of $l_{2r}$ with its respective constant. So in each step, only the linear terms will change between $q_1$ and $q_2$. Since each step only changes the quadratic terms, the decomposition takes the same amount of steps for each quadratic form, and therefore, the Witt rank is the same. \\

\textbf{Theorem 5.9 (Tsukiji)} \textit{If $q$ has Witt rank $\frac{n}{2} - c$ then $2^d$ is the sum of at most $2^{O(c)}$ full-rank quadratic characters.}\\

\textbf{Proof} Let $d$ = $l_1l_2 + ... + l_{2r-1}l_{2r} + l_0$ via its Witt decomposition. Since each of these terms are linearly independent, they form a basis for a subspace of linear forms. We may augment this set with $2c$ linearly independent forms $l_{2r+1}, ..., l_{n}$ in order to get a full basis for the linear functions of $n$ variables. Then $2^d = 2^{l_1l_2 + ... + l_{n-1}l_n}\cdot2^{l_{2r+1}l_{2r+2} + ... + l_{n-1}l_n}\cdot2^{l_0}$. The first term has clearly already been put into a Witt decomposition, and therefore has Witt rank $n/2$, or full Witt rank. The third term, $2^{l_0}$, contains only linears and constants, and thus can be combined with the first term without changing its Witt rank, by Theorem 5.8, which we will denote as $2^f$. The second term may be written as $\prod_{i=r+1}^{n/2} 2^{l_{2i-1}l{2i}}$. By Theorem 5.6, each term in the product may be written as a sum of four linear characters with respect to the new basis ${l_1, ..., l_n}$. We can rewrite the product as the product of $c$ sums of four linear characters, which will give us a sum of at most $2^{2c}$ linear characters. So we have $2^d = 2^f \sum_{i=1}^{2c} 2^{j_i}$, where $j_i$ is a linear character. Since adding these linear characters does not change the Witt rank, we may multiply the sum by $2^f$ to get the sum of $2c$ full rank quadratic characters. We conclude that any Witt rank $n/2-c$ quadratic character may be expressed as the sum of at most $2^{O(c)}$ full-rank quadratic characters. \\

\textbf{Corollary 5.9.1 (Tsukiji)} \textit{If $AND_n$ is the sum of $s$ arbitrary quadratic characters, $AND_n$ is also the sum of at most $s2^{O(\sqrt[]{n})}$ full-rank quadratic characters.} \\

\textbf{Proof} Suppose that $AND_n = \sum c_{\alpha}2^{\alpha}$, where the $\alpha$'s are quadratic forms and let $r$ be a random quadratic form. Then we have that $2^r AND_n = \pm AND_n = 2^r \sum c_{\alpha}2^{\alpha} = \sum c_{\alpha}2^{\alpha + r}.$We pick $c$ such that $n/2-c\sqrt[]{n}$ is less that $1/s$. Then, the probability that all terms have rank greater than $n/2-c\sqrt[]{n}$ is $(s-1)/s$, which is less than 1, and so there must be some quadratic character such that $r$ causes all of the $\alpha + r$ terms to have at least this Witt rank. By Theorem 5.9, each of these characters may be written as the sum of $2^{O(\sqrt[]{n})}$ full rank characters, and so we have $2^{O(\sqrt[]{n})}$ full rank characters. If this sum is equal to $-AND_n$, we multiply each character by 2 to produce $AND_n$, which clearly does not change the Witt rank of the forms. \\

Barrington conjectured that there is a strong correlation between Witt rank, support, and 2-weight, and that a bound on the 2-weight of a function on $n$ variables can be given in terms of the support of the function. The Tsukiji problem would then be a specific example of this conjecture, since $AND$ is a support 1 function. We present the conjecture and some related theorems, with a proof of the 2-weight-support trade-off for $n \leq 4$. \\

\textbf{Theorem 5.10} \textit{The support of a non-constant linear form is $2^{n-1}$.} \\

\textbf{Proof} We proceed by induction. For the base case, with $n = 1$, there are only two non-zero linear forms, $2^{x_1}$ and $2^{x_1+1}$, both of which have support 1. Assume that linear forms over  $n$ variables have support $2^{n-1}$. A linear form over $n+1$ variables has an $x_{n+1}$ term or it does not. If it doesn't, then it is a linear form over $n$ variables and has support $2^n-1$ by our inductive hypothesis. Over $n+1$ variables, each of the times that the $n$ variables yields a non-zero result would be repeated, once when the $n$th bit is zero, and once when it is one, so the support must be $2\cdot2^{n-1} = 2^{n}$. If it does have an $x_n$ term, then when $x_n$ = 0, the form is not affected. But when $x_n$ = 1, all of the 0's in the $n$ variables form become 1's, and vice verses. Since there were $2^{n}-2^{n-1}=2^{n-1}$ 0's, we now have support $2^{n-1} + 2^{n+1} - 2 = 2^{n}$. This closes the induction, and we conclude that this holds for all linear forms. \\

\textbf{Corollary 5.10.1} \textit{The support of a non-constant multilinear polynomial of degree d is at least $2^{n-d}$.} \\

\textbf{Proof} If n = d, it is clear that the support of the function must be at least one, since $2^{n-d} = 2^0$ in this case, a non-zero constant function. So the support must be at least $2^0 = 1$. Suppose that the minimum support for $n$ variables is $2^{n-d}$. We proceed by induction. A polynomial of degree $d$ over $n+1$ variables may be written as $q + rx_{n + 1}$, where $q$ and $r$ are polynomials over the first n variables, and each have degree at most $d$. If $x_{n+1}$ evaluates to 0, the number of non-zero solutions is the support of $q$ for $n$ variables, which, by our inductive assumption, is at least $2^{n-d}$. If $x_{n+1}$ evaluates to 1, then the number of solutions is the support of $r+q$, which is a polynomials of at most degree $d$, and therefore, again by our inductive assumption, is at least $2^{n-d}$. So the total support of the multilinear polynomial is $2^{n-d}+2^{n-d}=2^{(n+1)-d}$. By induction, this holds for all $n \in \mathbb{N}$, and since $d$ was arbitrary, it hold for all $d$ as well.  \\

The following is a correction of Theorem 6.14 in \cite{sindelar15}. \\

\textbf{Theorem 5.11} \textit{A family of quadratic forms with Witt rank $r$ over $n$ variables has $2^{2r}$ elements with support $2^{n-1} + 2^{n-r-1}$, $2^{2r}$ with support $2^{n-1} - 2^{n-r-1}$, and the rest have support $2^{n-1}$.} \\

\textbf{Proof} We first note that if there is a quadratic form $q_1$ with support $2^{n-1}-2^{n-r-1}$, then there is a quadratic form $q_1$ + 1 which will flip the bit output for any input, which shall give support $2^n - 2^{n-1}-2^{n-r-1} = 2^n + 2^{n-r-1}$. By a similar argument, if there is a quadratic form $q_2$ with support $2^{n-1} + 2^{n-r-1}$, there must be one of support $2^{n-1}-2^{n-r-1}$, namely $q_2$ + 1. 

Now, we proceed by induction. For the base case, with $r$ = 1, consider the form consisting of the term $l_1l_2$. The number of cases in which $l_1$ = 1 is $2^{n-1}$, and similarly for $l_2$ = 1. But because they are linearly independent, the number of times which their product is 1 must be $2^{n-2} = 2^{n-1}-2^{n-1-1}$. If the Witt decomposition is $l_1l_2 + l_0$ instead, than the support of $l_0$ is $2^{n-1}$, and since $l_0$ and $l_1l_2$ are linearly independent, then there are $2^{n-2}*2^{n-1}*2^{-n}$ = $2^{n-3}$ places where they both must be 1.  Thus the total support is $2^{n−1}+2^{n−2}−2*2^{n−3} = 2^{n−1}$. Now, if we consider the forms $l_1l_2 + \sum_{i = 1}^{2} a_il_i + c$, with $a_i \in \mathbb{Z}_2$, we now have a form in the same family, meaning they have the same support, which we may write as $m_1m_2 + c$, where $m_1$, $m_2$ are other linear forms and some $c \in \mathbb{Z}_2$. We have therefore shown that for a family with Witt rank 1 and decomposition $l_1l_2$, $l_1l_2 + \sum_{i = 1}^{2} a_il_i$ has support $2^{n−1}−2^{n−r−1}$, $l_1l_2 + \sum_{i=1}^{2} a_il_i + 1$ has support $2^{n−1}+ 2^{n−r−1}$, and $l_1l_2 + l_0$ has support $2^{n−1}$.

Given that any family of rank $r$ has exactly $2^{2r}$ elements of the form $l_1l_2 + ... + l_{2r−1}l_{2r} + \sum_{i=1}^{2r} a_il_i$ that have support $2^{n−1} − 2^{n−r−1}$, consider a decomposition of rank $r$ + 1 with $l_1l_2 + ... + l_{2r−1}l_{2r}+l_{2r+1}l_{2r+2} + \sum_{i=1}^{2r+2} a_il_i$. The $l_{2r+1}l_{2r+2}$ term has support $2^{n−2}$ by the base case. Since they are linearly independent then both $l_1l_2 + ... + l_{2r−1}l_{2r}$ and $l_{2r+1}l_{2r+2}$ evaluate to 1 on $2^{n−2} \cdot (2^{n−1}−2^{n−r−1}) \cdot 2^{−n} = 2^{n−2} \cdot 2^{−1}−2^{n−2} \cdot 2^{r−1} = 2^{n−3} − 2^{n−r−3}$ different inputs. Now, adding any combination of linear terms $\sum_{i=1}^{2r+2} a_il_i$ will again transform the form into another in the same family, and with the same support. Since there are $2^{2(r+1)}$ ways to do so, there must be $2^{2(r+1)}$ forms of this support. So the support of these forms is then $2^{n−1} − 2^{n−r−1} + 2^{n−2−2} \cdot (2^{n−3}−2^{n−r−3}) = 2^{n−1} − 2^{n−r−1} + 2^{n−r−2} = 2^{n−1} − 2^{n−(r+1)−1}$. If we add the constant 1 to any of these forms, we shall flip each of the bits of evaluation, and thus we also have $2^{2(r+1)}$ quadratic forms with support $2^{n−1} + 2^{n−(r+1)−1}$. 

Finally, let the decomposition be $l_1l_2 + ... + l_{2r−1}l_{2r} + l_{2r+1}l_{2r+2} + l_0$. By the inductive hypothesis, we have that $l_1l_2 + ... + l_{2r−1}l_{2r} + l_0$ has support $2^{n−1}$. The $l_{2r+1}l_{2r+2}$ term has support $2^{n−2}$ by the base case. Since they are linearly independent, both $l_1l_2 + ... + l_{2r−1}l_{2r} + l_0$ and $l_{2r+1}l_{2r+2}$ are 1 on $2^{n−1} \cdot 2^{n−2} \cdot 2^{−n} = 2^{n−3}$ different inputs. So the support is then $2^{n−1} + 2^{n−2} − 2 \cdot 2^{n−3} = 2^{n−1}$. Thus we have shown for a family of quadratic forms with Witt rank $r$ + 1 there are $2^{2r}$ elements of support $2^{n−1} + 2^{n−r−1}$ of the form $l_1l_2 + ... + l_{2r−1}l_{2r} + l_{2r+1}l_{2r+2} + 1$, $2^{2r}$ of support $2^{n−1} − 2^{n−r−1}$ of the form $l_1l_2 + ... + l_{2r−1}l_{2r} + l_{2r+1}l_{2r+2}$, and the rest have support $2^{n−1}$, and therefore, this holds for families of quadratic character of any Witt rank $r$ by induction. \\

\textbf{Conjecture 5.3 (Barrington)} \textit{If any function has 2-weight $w \geq 1$ and support $s$, then $w^2s \geq 2^{n}$.} \\ 

\textbf{Theorem 5.12} \textit{The conjectured 2-weight support trade-off is true for $w \leq 4$.} \\

\textbf{Proof} For the $w= 1$ case, a quadratic character $2^{q(x)}$ is nonzero for any form $q(x)$, and so a function must be nonzero everywhere. This function always has support $2^n$. 

For the $w$ = 2 case, we have functions of the form $2^{p(x)} + 2^{q(x)}$. These may be factored as $2^{p(x)}(1 + 2^{q'(x)})$, which evaluates to 0 if and only if $q'(x) = 0$. The support any quadratic form is at most $2^{n-1} + 2^{n-r-1}$, where $r$ is its Witt rank (by Theorem 5.11), and so there must be at least $2^{n-1} - 2^{n-r-1}$ zero evaluations of $q'(x)$. Now, if $q'(x)$ has a Witt rank of 1, then it is a linear form and has support $2^{n-1}$. Otherwise, the form has at least $2^{n−1}−2^{n−r−1}$ non-zero evaluations, where $r$ $\geq$ 1. So $s \geq 2^{n−1}−2^{n−r−1} \geq 2^{n−1}−2^{n−2}= 2^{n−2}$, and thus, $w^2s= 4s \geq 2^n$.

For the $w$ = 3 case, the function is a sum of a 2-weight 2 function and a 2-weight 1 function. As we saw in case 1, the support of the 2-weight 1 function is $2^n$, and so the support of the 2-weight 3 function in question is nonzero when one function is and when one isn't, so the support is at least the symmetric difference of the two functions' support sets. So, because the 2-weight 1 function has support $2^n$, and the 2-weight 2 function has support at most $2^{n-1} + 2^{n-2}$ = $(\frac{3}{4})2^n$, the difference has at least support $(\frac{1}{4})2^n$, and so $s \geq (\frac{1}{4})2^n$, and we have that $3^2s \geq (\frac{9}{4})2^n \geq 2^n$. 

For the $w$ = 4 case, it is clear that we are dealing with the sum of two 2-weight 2 functions, $f_1$ and $f_2$. We write $f_1 = 2^{p_1}+ 2^{q'_1} = 2^{p_1}(1 + 2^{q_1})$ and $f_2 = 2^{p_2} + 2^{q'_2} = 2^{p_2}(1 + 2^{q_2})$, in a similar fashion to the factor of a 2-weight 2 function in case 2. If $q_1 \neq q_2$, then the function is nonzero whenever $q_1(x) \neq q_2(x)$, since this will cause one of the $f_i$'s to evaluate to zero and the other to have a nonzero evaluation. The support of $q_1 + q_2$ is at least $2^{n−1}−2^{n−r−1} \geq 2^{n−1}−2^{n−2} \geq 2^{n−2}$and so they must disagree in at least $2^{n−2}$ places. So, the function has support $s$ $\geq 2^{n−2}$, and we have that $4^2s \geq 2^{n+2} \geq 2^n$.

Now, if $q_1 = q_2$, we may write the function as $2^{p_1}(1 + 2^q) + 2^{p_2}(1 + 2^q) = (2^{p_1}+ 2^{p_2})(1 + 2^q)$. This function evaluates to a nonzero value only if $p_1(x) = p_2(x)$ and $q(x) \neq 1$. So it is nonzero when 
If $q_1 = q_2$, then we can write the function as  $(q+ 1)(p_1+p_2+ 1) = 1$. This product cannot be the zero polynomial, since otherwise, the function is identically zero and does not have weight 4. It is a multilinear polynomial of degree at most 4, and so its support $s$ is at least $2^{n-4}$. So we have that $4^2s \geq 2^n$. 

\section{Previous Work}

As inspiration for this work, we took many of the above results from Sindelar \cite{sindelar15}, and hope to expand on the analytic results he was able to obtain. Here, we shall discuss briefly the experiments that he performed, and the results he reported. 

Sindelar wished to find a property of quadratic characters that would satisfy similar conditions to those of linear characters, namely that a random quadratic character possesses said property with low probability, but the product of a random character with the $AND$ function must have. Sindelar began by constructing a database of all possible functions from $\mathbb{Z}_2^4$ to $\mathbb{Z}_3$, and saved information about the 2-weight functions and the sums to produce them. Using different bases, he generated datasets to see the effect of constraining quadratic characters to certain sets that share specific properties, in order to understand the effect that these properties have on the weight of $AND$.

\subsection{Observed 2-weight of $AND_4$}

Sindelar wanted to experimentally observe the 2-weight of $AND_4$. To do this, he built a table of all functions from $(\mathbb{Z}_2)^4$ to $\mathbb{Z}_3$, and treated each function as a node in a massive graph, where the edges corresponded to the addition of quadratic characters. He then performed breadth-first search from the zero function, and found that the minimal number of quadratic characters need to sum to the $AND$ function on 4 variables was 4, consistent with the previously known fact that the 2-weight of $AND_4$ is 4. One such sum of these characters was $2^0 + 2^{x_1x_2 + 1} + 2^{x_3x_4+1} + 2^{x_1x_2+x_3x_4}$, which you may verify for all settings of the $x_i$'s.

\section{Experiments and Results}

In order to make progress on the Constant Degree Hypothesis, we decided to investigate functions from $(\mathbb{Z}_2)^6$ to $\mathbb{Z}_2$, and therefore, we hope to find that the 2-weight of $AND_6 = 2^3 = 8$. Unfortunately, we could not implement the same graph search that Sindelar used to observe the 2-weight of $AND_4$, since exhaustively generating all functions of a given 2-weight over 6 variables quickly becomes intractable as the 2-weight increases. So instead, we began by randomly sampling functions of small 2-weight (2, 3, 4, 6, ...) in order to get a feel for their support distributions. 

\subsection{Random Sampling Functions from $(\mathbb{Z}_2)^6$ to $\mathbb{Z}_3$}

Our first approach was to randomly generate functions of small 2-weight, in order to get a rough sketch of their support distributions. Moreover, we believed this sampling will serve as corroborative evidence for the Constant Degree Hypothesis, in that we didn't expect to see any functions of n-weight less than 8 of support 1. Below are the support distributions of one hundred thousand randomly distributed functions of various 2-weights.

\begin{center}
Table 3: Supports of Randomly Sampled Functions of 2-weight 2
\break 
\begin{tabular}{ |c|c|c|c|c|c|c|c| }

\hline
support: & 16 & 24 & 28 & 32 & 36 & 40 & 48 \\
\hline
functions: & 65 & 7084 & 21111 & 43488 & 21175 & 7014 & 63 \\
\hline

\end{tabular}
\end{center}

\hfill

\begin{center}
Table 4: Supports of Randomly Sampled Functions of 2-weight 3
\break \hfill
\begin{tabular}{ |c|c|c|c|c|c|c|c|c|c|c|c|c|c|c|c|}

\hline
support: & 32 & 34 & 36 & 38 & 40 & 42 & 44 & 46 & 48 & 50 & 52 & 54 & 56 & 58 & 60 \\
\hline
functions: & 7 & 10 & 93 & 503 & 1934 & 5077 & 11550 & 18467 & 23198 & 19784 & 12584 & 5219 & 1356 & 205 & 13 \\
\hline

\end{tabular}
\end{center}

\hfill


\begin{center}
Table 5: Supports of Randomly Sampled Functions of 2-weight 4
\break \hfill
\begin{tabular}{ |c|c|c|c|c|c|c|c|c|c|c|c|c|c|c|c|c| }

\hline
support: & 20 & 23 & 24 & 25 & 26 & 27 & 28 & 29 & 30 & 31 & 32 \\
\hline
functions: & 1 & 1 & 1 & 6 & 6 & 24 & 99 & 188 & 405 & 764 & 1397 \\
\hline \hline
33 & 34 & 35 & 36 & 37 & 38 & 39 & 40 & 41 & 42 & 43 & 44 \\
\hline
2139 & 2979 & 4242 & 5737 & 7430 & 8995 & 9910 & 10460 & 10265 & 8937 & 7488 & 6107 \\
\hline \hline
45 & 46 & 47 & 48 & 49 & 50 & 51 & 52 & 52 & 54 & 55 & 57 \\
\hline
4714 & 3191 & 2071 & 1209 & 687 & 316 & 134 & 55 & 16 & 23 & 2 & 1 \\
\hline

\end{tabular}
\end{center}

\hfill

\begin{center}
Table 6: Supports of Randomly Sampled Functions of 2-weight 6
\break \hfill
\begin{tabular}{ |c|c|c|c|c|c|c|c|c|c|c| }

\hline
support: & 25 & 27 & 28 & 29 & 30 & 31 & 32 & 33 & 34 & 35 \\
\hline
functions: & 3 & 4 & 24 & 33 & 81 & 203 & 383 & 720 & 1162 & 1971 \\
\hline \hline
36 & 37 & 38 & 39 & 40 & 41 & 42 & 43 & 44 & 45 & 46 \\
\hline
2925 & 4406 & 5754 & 7459 & 8986 & 10214 & 10400 & 10172 & 9225 & 7904 & 6067 \\
\hline \hline
47 & 48 & 49 & 50 & 51 & 52 & 53 & 54 & 55 & 56 & 57 \\
\hline
4603 & 3148 & 1897 & 1202 & 574 & 306 & 104 & 50 & 16 & 3 & 1 \\
\hline

\end{tabular}
\end{center}

\hfill

From these tables, we can see that finding 2-weight 2, 3, 4, or 6 functions of very small or very large support is relatively rare. Now, since the $AND$ function has support 1 for all $n$ (when $x_i$ = 1 for all $i \in \{1, 2, ..., n\}$), we might not randomly generate the $AND_6$ in a feasible amount of time, if it were to have such a 2-weight. This lends support to the belief that the $AND_6$ function may not be computable by a function of the above sampled weights.

\subsection{Characterizing Functions of 2-weight 3}

Now, it is computationally impractical to exhaustively generate all functions of 2-weight 3, as there are some $(2^{22})^3$ functions of this kind. Therefore, creating a graph with this many nodes in it and running breadth-first search on it is intractable. 

To mitigate this, we take advantage of the fact that any 2-weight 3 function, $f = 2^p + 2^q + w^r$, for quadratic forms $p$, $q$, and $r$, may be written as $f' = 2^p(1 + 2^s + 2^t)$, for different quadratic forms $s$ and $t$. Clearly, $f'$ and $f$ have the same support, as it is the same function. By change of basis, we say that $f'$ has the same support as the function $g = 1 + 2^u + 2^v$, where $u$ is in Witt Normal Form. In our case, with $n$ = 6, we know that $u$ must be one of 

\begin{center}
$\{0, 1, x_1, x_1 + 1, x_1x_2, x_1x_2 + 1, x_1x_2 + x_3, x_1x_2 + x_3 + 1, x_1x_2 + x_3x_4, x_1x_2 + x_3x_4 + 1, x_1x_2 + x_3x_4 + x_5, x_1x_2 + x_3x_4 + x_5 + 1, x_1x_2 + x_3x_4 + x_5x_6, x_1x_2 + x_3x_4 + x_5x_6 + 1\}$,
\end{center}

and so, by exhaustively searching over the 14 choices for $u$ and $2^{22}$ choices of $v$, we will get the exact distribution for all functions of 2-weight 3. In Table 7, we place the weighted sum (by occurrence of family of Witt normal form) of the support distribution of all those 2-weight 3 functions we were able to identify using this technique, which has been normalized to sum to one hundred thousand, to compare with the randomly sampled distributions above.

\hfill

\begin{center}
Table 7: Support Distribution of all 2-weight 3 Functions (Normalized to sum to 100,000)
\break \hfill
\begin{tabular}{ |c|c|c|c|c|c|c|c|c|c|c| }

\hline
support & 0 & 16 & 24 & 28 & 32 & 34 & 36 & 38 & 40 & $-$ \\
\hline
functions & 0 & 0 & 0 & 0 & 4 & 10 & 97 & 489 & 1826 & $-$ \\
\hline \hline
42 & 44 & 46 & 48 & 50 & 52 & 54 & 56 & 58 & 60 & 64 \\
\hline
5051 & 11522 & 18497 & 23319 & 19797 & 12662 & 5134 & 1379 & 194 & 20 & 0 \\ 
\hline

\end{tabular}
\end{center}

\hfill

From this table, it is clear that the distribution is clustered around supports near 48, which is the expected behavior of a randomly sampled function of 2-weight 3 by the following argument: Consider an arbitrary 2-weight 3 function, $2^q + 2^r + 2^s$. We can define 8 families of functions by replacing any of $p$, $q$, or $r$ by its inverse, $p+1$, $q+1$, or $r+1$, in any combination. Looking at the function on arbitrary input, it will evaluate to 1+1+1, 1+1+2, 1+2+1, 2+1+1, 1+2+2, 2+1+2, 2+2+1, or 2+2+2, the order and occurrence of these determined by the quadratic forms $q$, $r$, and $s$ and the family to which the function belongs. Exactly two of these function evaluations are equivalent to zero modulo 3, and therefore the average support of an arbitrary function of 2-weight 3 is 48, since on average, a 2 weight 3 function will evaluate to 0 on a quarter of its inputs.

To get a better idea of the true distribution of all 2 weight 3 functions, we randomly sampled 10 million 2-weight 3 functions, and included a binary table of their 1's and 2's vectors, given below in Table 9. The x-axis represents the number of 1's given by a function, and the y-axis represents the number of 2's. Instead of including their true distribution, we mark a 1 in the positions where a function has evaluated to this support in one or more instances, and otherwise, mark a 0. The entry in the top left-hand corner represents the function evaluating to zero 1's and 2's, and the entry in the bottom right-hand corner represents the function evaluating to sixty-four 1's and sixty-four 2's. Accordingly, the lower right triangular portion of this distribution is identically zero, since there cannot be functions that evaluate to more than sixty-four 1's and 2's combined.

We may think of this distribution as members of the previous distribution, which are functions of the form $1 + 2^u + 2^v$, which have been multiplied by a random quadratic character. This multiplication keeps the support the same, but may swap the number of 1's and 2's in the function's evaluation, causing the entry to move northeast or southwest in the table while keeping the same support. As we saw in Table 7, no 2-weight 3 function has odd support, which is also evident in the entries of this table. 

\pagebreak

\begin{center}

Table 8: Binary Support Distribution of Randomly Sampled Functions of 2-weight 3

{\tiny 

\begin{longtable*}{|c|}

\hline

00000000000000000000000000000000000000000000001000001000000000000\\
00000000000000000000000000000000000000000000000000000000000000000\\
00000000000000000000000000000000000000000000000000000000000000000\\
00000000000000000000000000000000000000000000000000000000000000000\\
00000000000000000000000000000000000000000000000000000000000000000\\
00000000000000000000000000000000000000000000000000000000000000000\\
00000000000000000000000000000010100010101000000000000000000000000\\
00000000000000000000000000000101010101010100000000000000000000000\\
00000000000000000000000000101010101010101010101000000000000000000\\
00000000000000000000000101010101010101010101010000000000000000000\\
00000000000000000000001010101010101010101010100000000000000000000\\
00000000000000000000010101010101010101010101010000000000000000000\\
00000000000000000000101010101010101010101010100000000000000000000\\
00000000000000000001010101010101010101010101010000000000000000000\\
00000000000000100010101010101010101010101010100000000000000000000\\
00000000000001000101010101010101010101010101010000000000000000000\\
00000000000010001010101010101010101010101010000000000000000000000\\
00000000000100010101010101010101010101010000000000000000000000000\\
00000000001000101010101010101010101010101000000000000000000000000\\
00000000000001010101010101010101010101010000000000000000000000000\\
00000000000010101010101010101010101010101000000000000000000000000\\
00000000000101010101010101010101010101010000000000000000000000000\\
00000000001010101010101010101010101010100000000000000000000000000\\
00000000010101010101010101010101010101000000000000000000000000000\\
00000000101010101010101010101010101010000000000000000000000000000\\
00000000010101010101010101010101010100000000000000000000000000000\\
00000000101010101010101010101010101000000000000000000000000000000\\
00000000010101010101010101010101010000000000000000000000000000000\\
00000000101010101010101010101010100010000000000000000000000000000\\
00000001010101010101010101010101000000000000000000000000000000000\\
00000010101010101010101010101010000000000000000000000000000000000\\
00000001010101010101010101010100000000000000000000000000000000000\\
00000010101010101010101010101000100000000000000000000000000000000\\
00000001010101010101010101010000000000000000000000000000000000000\\
00000010101010101010101010100000000000000000000000000000000000000\\
00000001010101010101010101000000000000000000000000000000000000000\\
00000010101010101010101010001000000000000000000000000000000000000\\
00000001010101010101010100000000000000000000000000000000000000000\\
00000010101010101010101000000000000000000000000000000000000000000\\
00000001010101010101010000000000000000000000000000000000000000000\\
00000000101010101010100010000000000000000000000000000000000000000\\
00000001010101010000000000000000000000000000000000000000000000000\\
00000000101010100000000000000000000000000000000000000000000000000\\
00000001010101010000000000000000000000000000000000000000000000000\\
00000000001010100000000000000000000000000000000000000000000000000\\
00000000010101000000000000000000000000000000000000000000000000000\\
00000000000000100000000000000000000000000000000000000000000000000\\
00000000000000000000000000000000000000000000000000000000000000000\\
10000000000000001000000000000000000000000000000000000000000000000\\
00000000000000000000000000000000000000000000000000000000000000000\\
10000000000000000000000000000000000000000000000000000000000000000\\
00000000000000000000000000000000000000000000000000000000000000000\\
00000000000000000000000000000000000000000000000000000000000000000\\
00000000000000000000000000000000000000000000000000000000000000000\\
00000000000000000000000000000000000000000000000000000000000000000\\
00000000000000000000000000000000000000000000000000000000000000000\\
00000000000000000000000000000000000000000000000000000000000000000\\
00000000000000000000000000000000000000000000000000000000000000000\\
00000000000000000000000000000000000000000000000000000000000000000\\
00000000000000000000000000000000000000000000000000000000000000000\\
00000000000000000000000000000000000000000000000000000000000000000\\
00000000000000000000000000000000000000000000000000000000000000000\\
00000000000000000000000000000000000000000000000000000000000000000\\
00000000000000000000000000000000000000000000000000000000000000000\\
00000000000000000000000000000000000000000000000000000000000000000\\

\hline

\end{longtable*}}

\end{center}

In order sum to the $AND_6$ function, two 2-weight 3 functions must have opposing 1's and 2's vectors, in that both must have support 64, and where one evaluates to a 1, the other must evaluate to a 2, except for on a single input, where both functions have the value 2. In this way, the total 2-weight 6 function evaluates to 0 on every input except for one, on which it evaluates to 1. 

No two functions in our randomly sampled distribution have this property, and so no two of these may sum to the $AND_6$ function. However, we do not yet know if our randomly sampled distribution or our Witt Normal Form distribution correctly models the true distribution. For now, our best lower bound on the $AND_6$ function is 2-weight 5, since we know from Sindelar's work that no two 2-weight 2 functions have the property detailed above. 

We include a plot of the distribution of supports of all functions of 2-weight 3 for further inspection, again normalized to sum to one hundred thousand. The distribution roughly outlines a standard normal curve, and we can expect it to approach the normal distribution as we let the 2-weight of the functions we are generating go to infinity. There are but a few functions of support less than 24, which turn out to be trivial functions of 2-weight 1 or 2. This is useful data in support of Conjecture 5.2, which we may be able to extrapolate to a new lower bound on the $AND_6$ function.

\begin{center}

\includegraphics[scale=0.75]{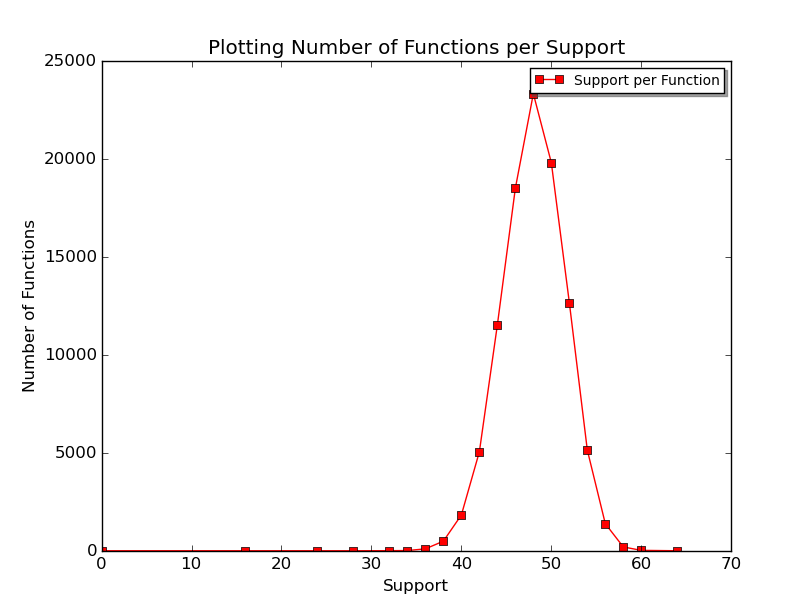}

\end{center}

\section{Conclusions}

We were not able to obtain a strict lower bound on the 2-weight of $AND_6$. But, the current work on functions of 2-weight 3 is promising in that we were not able to find any Witt Normal Form 2-weight 3 functions with total support that have opposite 1's and 2's vectors, except in one case where both evaluate to 2, in which case their sum would be the $AND_6$ function. Now, if this holds for the true distribution of all 2-weight 3 functions, we have a new lower bound of 2-weight 7. Further, if we can extend this technique to show that a 2-weight 3 cannot differ in exactly one place from a 2 weight 4 in the way mentioned above, then we may conclude that the $AND_6$ function must have 2-weight 8. We then hope to generalize this lower bound in order to prove Conjecture 5.2, therefore proving the Constant Degree Hypothesis for characters of degree 2 over the field $\mathbb{Z}_3$. 

\pagebreak

\end{document}